\documentclass[%
 reprint,
bibnotes,
 amsmath,amssymb,
 aps,
letterpaper
]{revtex4-2}

\usepackage{graphicx}
\usepackage{dcolumn}
\usepackage{bm}
\usepackage[dvipsnames]{xcolor}
\usepackage{hyperref}

\newcommand{\dip}[1]{\times 10^{#1}}

\begin{document}

\title{A compact ultracold strontium experiment}

\author{Mathieu Bertrand}
\email{Present address: Thales Alenia Space, Toulouse, France}
\author{Torben P{\"o}pplau}
\email{Present address: Trumpf SE + Co. KG, Ditzingen, Germany}
\author{Luca Beguin}
\email{Present address: Pasqual SAS, Palaiseau, France}
\author{Romain Long}
\author{Jakob Reichel}
\email{jakob.reichel@ens.fr}
\affiliation{
Laboratoire Kastler Brossel, ENS-Universit{\'e} PSL, CNRS, Sorbonne Universit\'e, Coll{\`e}ge de France, 24 rue Lhomond, 75005 Paris, France}
\date{\today}

\begin{abstract}
We describe a compact and simplified experimental setup for the production of ultracold strontium atoms. Thanks to its low-power oven and coil-free Zeeman slower design, the vacuum system has an overall length of only 40\,cm including oven, Zeeman slower and all-glass science cell. The oven has a power consumption of 14\,W at $460^\circ$C and produces a collimated flux of $3\cdot 10^{10}\mbox{at}\mbox{s}^{-1}\mbox{cm}^{-2}$ at this temperature. The laser system employs master-slave laser architecture throughout and requires neither second-harmonic generation nor tapered amplifiers, reducing cost and complexity. The system has been in continuous use for more than a year.
\end{abstract}

\maketitle

\section{Introduction}
\label{sec:intro}

Earth alkali-like atoms have expanded the gamut of ultracold atom experiments.
The existence of long-lived metastable states, combined with ``magical''
trapping wavelengths \cite{Katori03}, have enabled optical lattice clocks
\cite{Derevianko11,Ludlow15}, which are approaching and even surpassing
\cite{Aeppli24} the $10^{-18}$ systematic relative uncertainty level. Their
versatile internal structure has also led to the realization of SU(N)-symmetric
quantum many-body systems \cite{Gorshkov10,Taie10,Zhang14,Scazza14,Zhang20},
synthetic dimensions and artificial gauge fields \cite{Mancini15,Livi16} and has
enabled continous Bose-Einstein condensation \cite{Chen22}. In optical lattices
and optical tweezers, earth alkali-like atoms open new roads for quantum
simulation and quantum computing
\cite{Schine22,Ma22,Cao24,Finkelstein24,Buob24,Senoo26,Tao26}. Alkaline-earth
Rydberg atoms can be trapped in the same tweezers as the ground state atom
\cite{Wilson22}, and this has been extended to circular Rydberg atoms
\cite{Hoelzl24}. The combination of optical clock and tweezers technologies
\cite{Norcia19, Madjarov19,Young20} is an inspirational new platform for
metrology applications, where systematic shifts can be evaluated in parallel,
and where the same entanglement production methods that serve for quantum
simulation and computing are readily available for quantum-enhanced measurement
and sensing. Combination of alkali earth atoms with Cavity QED offers offers
several perspectives. A continuous superradiant laser, which would realize an
active optical clock, is the subject of a long-standing effort
\cite{Norcia16a,Laske19,Schaffer20,Gogyan20,Kazakov21,Kristensen23,Cline25}.
Cavities have also been used to simulate dynamical phases of BCS superconductors
\cite{Young24}, and enable spin squeezing, which is particularly topical for the
optical clock transitions in alkaline-earth atoms. It has been first been
demonstrated for Yb \cite{Braverman19,Pedrozo-Penafiel20} and has recently been
used in Sr in a clock measurement beyond the standard quantum limit at a
fractional frequency instability level of $8.0\cdot 10^{-17}/\sqrt{\tau}$ level
\cite{Yang25}. 

However, production of ultracold alkaline-earth-like atoms remains challenging
and puts a high entrance barrier before alkaline-earth experiments. Efforts have
been made to reduce the size and complexity of Sr and Yb experiments in
particular in the context of transportable optical clocks \cite{Schioppo12,Kale22,Kwon23}. 
Here we describe a setup for ultracold Sr production that has a compact vacuum system 
(40\,cm total length), no Zeeman coils, and relies on extended-cavity diode lasers (ECDLs) and 
injected diodes exclusively, requiring neither tapered amplifiers nor second-harmonic generation.

\section{Vacuum system}

\begin{figure}[htb]
\includegraphics[width=0.9\columnwidth]{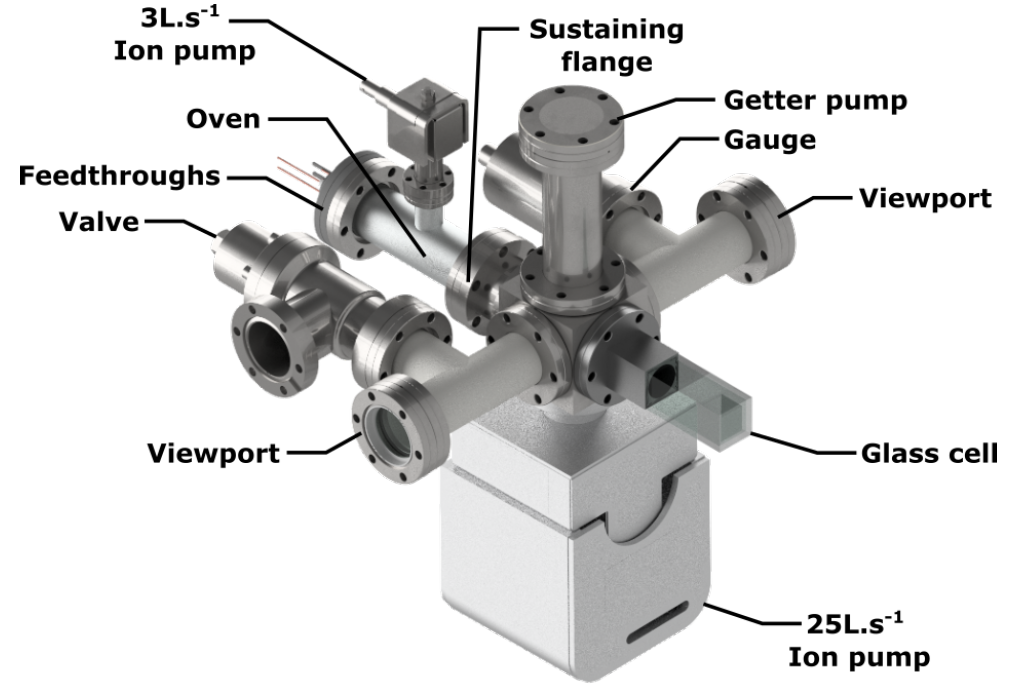}\\
\includegraphics[width=0.9\columnwidth]{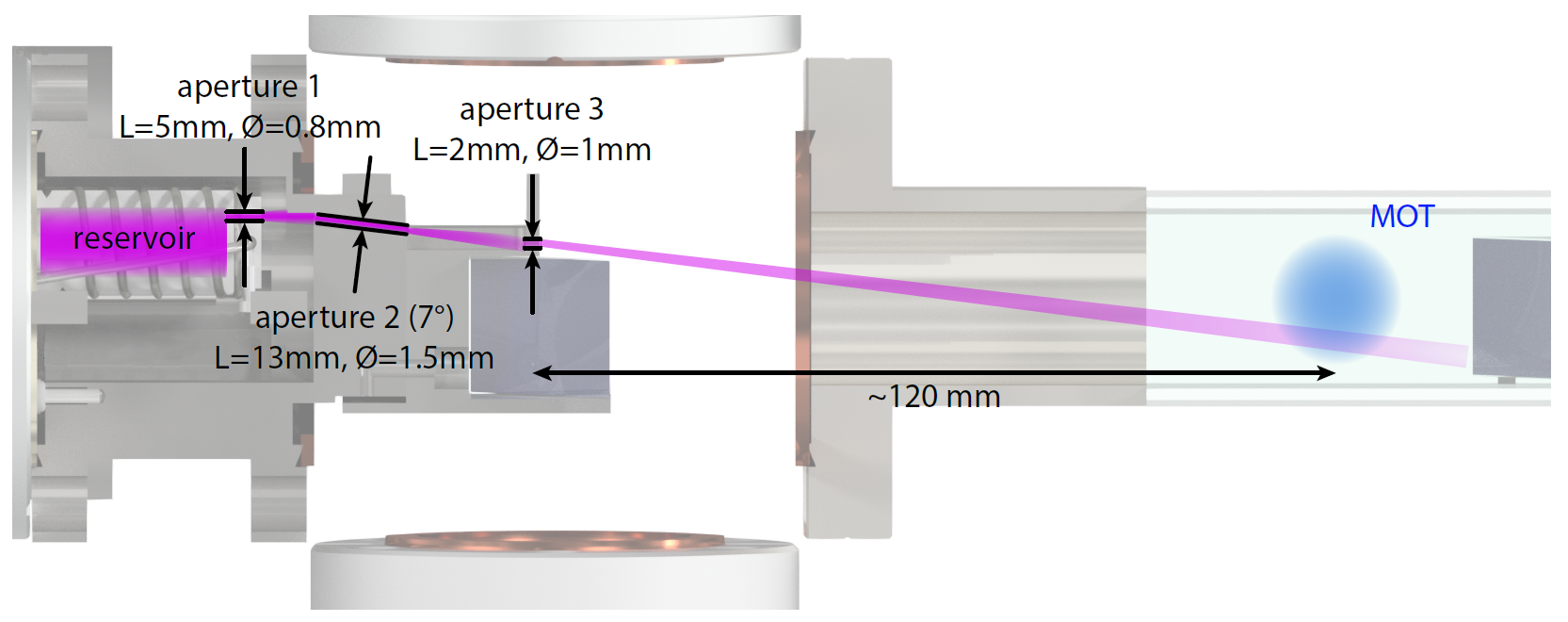}
\caption{\label{fig:vacuum} (a) Global view of the vacuum system. The total length from the 
feedthrough to the glass cell is 40\,cm. (b) Sectinal view showing the atomic beam path, with the
six-way cross removed for clarity.  }
\end{figure}

Fig.~\ref{fig:vacuum}(a) shows an overview of the vacuum system. Its total
length is 40\,cm, measured in the atomic beam direction, from the electrial
feedthroughs to the end of the glass cell. The components of the system are
attached to a six-way 40CF cross in the center. The atomic beam is generated by
an oven contained in the tee on the left, and enters the all-glass science
chamber on the right. The largest component is a $25\,\mbox{l}/\mbox{s}$ ion
pump, which pumps the science chamber and also serves as the base of the system.
Fig.~\ref{fig:vacuum}(b) shows the atomic beam path.

Main design features of this apparatus are a simple and low-power oven, a
short distance from oven to science chamber which crucially increases atomic
flux in the MOT capture region, and the reuse of the MOT coil field for Zeeman
slowing, eliminating the need for Zeeman coils. The short length is enabled by a
low-cost glass cell glued to a titanium flange. In the following, we describe these
components in more detail.


\subsection{Oven}
\label{sec:oven}
In contrast to alkalis such as Rb and Cs, which have low melting points (39 and
28$^\circ$C) and high vapor pressure at room temperature, alkaline earth-like
elements have much higher melting points in the 650 to 850$^\circ$C range and
need to be heated to release any significant amount of atoms into the vapor
phase. Ovens for these elements typically operate at $400\ldots 500^\circ$C for
Yb, and $450\ldots 550^\circ$C for Sr, to produce sufficient atom flux for
trapping. These ovens are one main source of experimental complexity, the two
main issues being the thermal management and the output nozzle. Designs aiming
for maximised brightness have a tendency to favor clogging of the nozzle -- an
experimentalist's spectre because of the long time constant associated with
opening the vacuum chamber for repair. Recently demonstrated miniature,
laser-heated or resistively heated dispensers \cite{Pick25} may provide an
attractive alternative to ovens in some situations, but their non-directional
emission can be a concern when nearby surfaces should not be exposed to the
vapor, which is the case in a cavity experiment such as ours.

\begin{figure*}[htb]
\includegraphics[width=0.9\columnwidth]{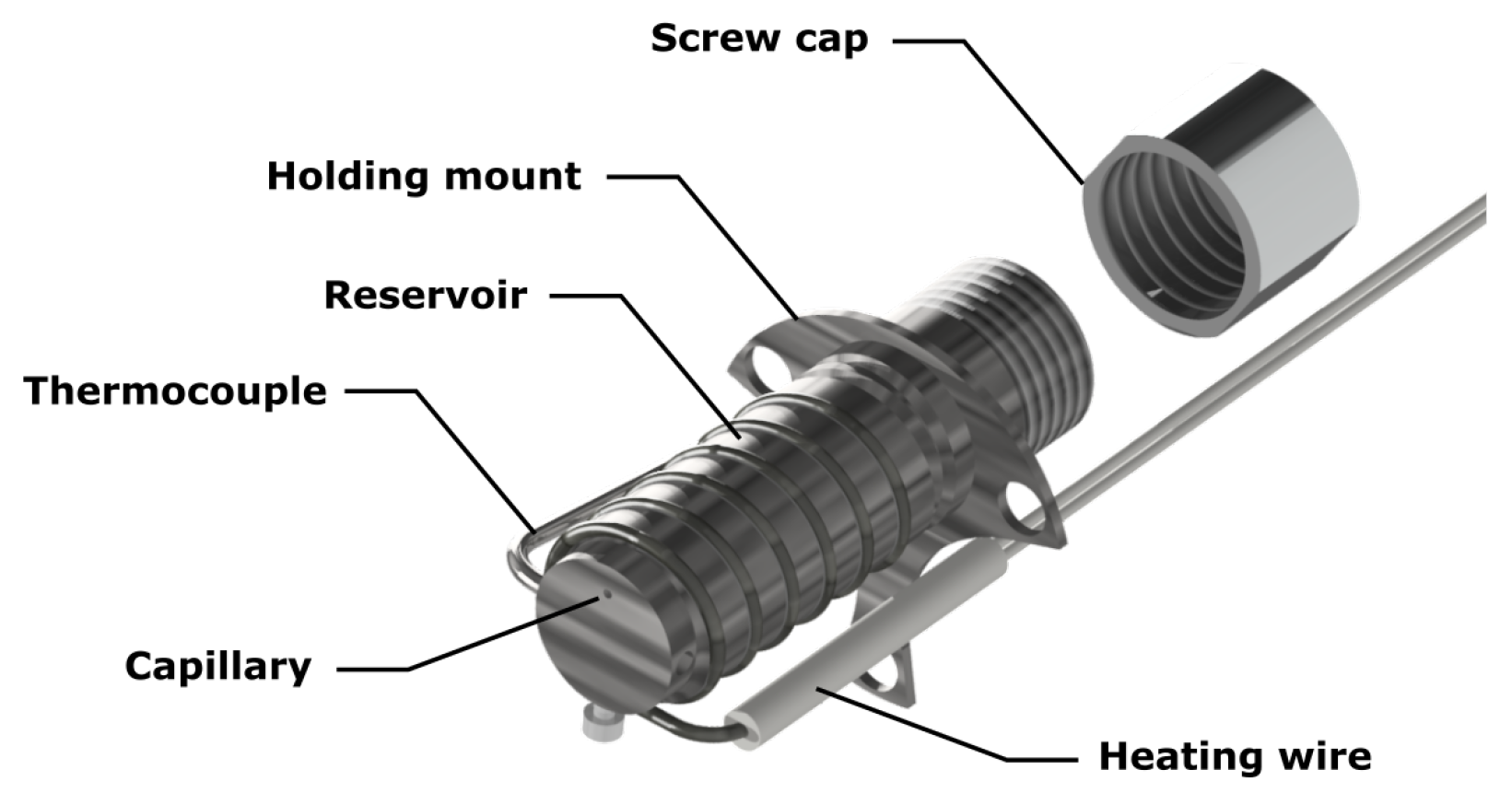}
\includegraphics[width=0.9\columnwidth]{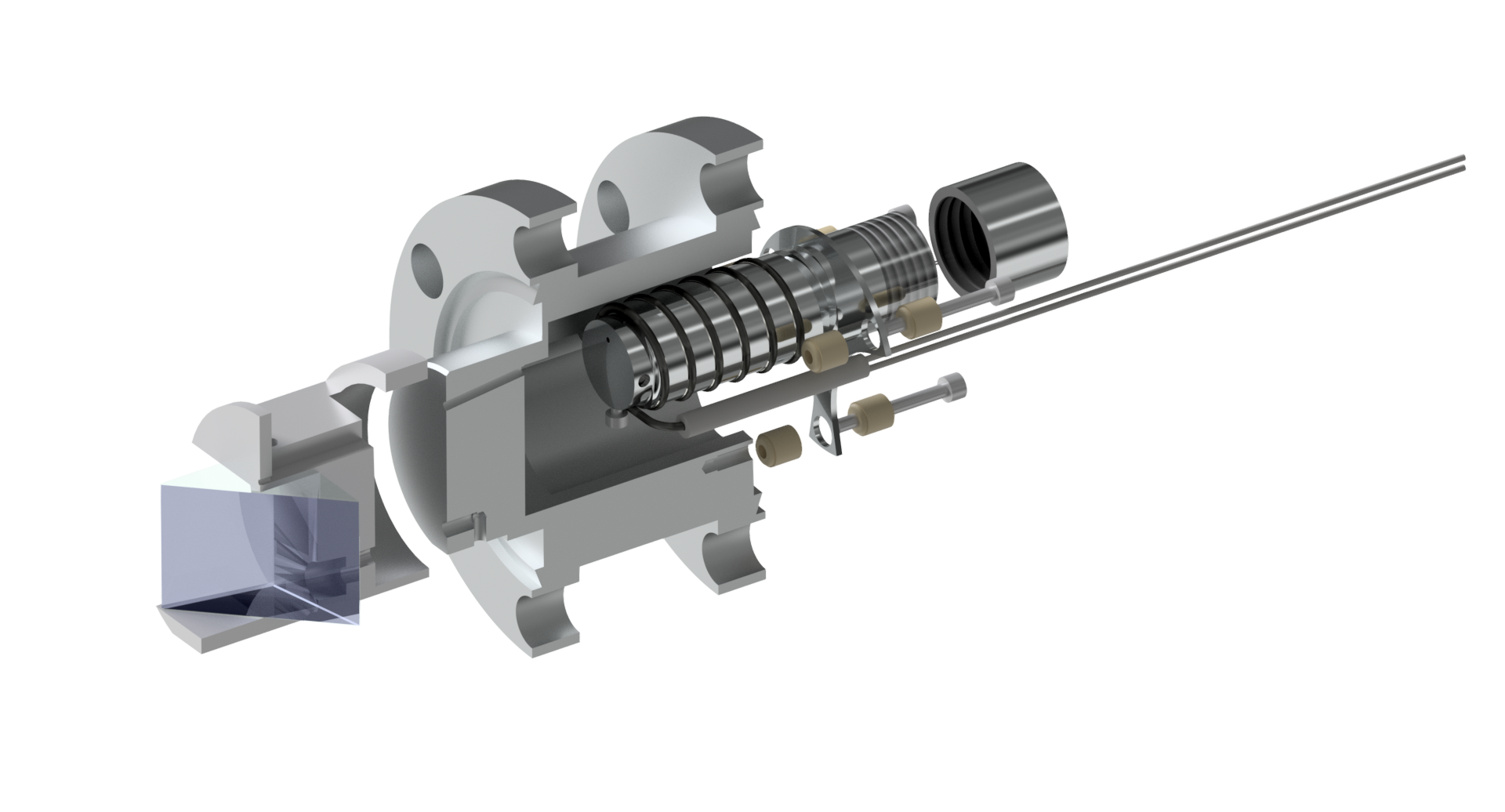}
\caption{\label{fig:oven} (a) CAD view of the oven. (b) In-situ view of the oven assembly in the 
sustaining flange. (Thermocouple not shown.) Low thermal conduction between the oven and the the flange 
is ensured by the reduced thickness of the holding plate and the nonmetallic screw sleeves and spacers.}
\end{figure*}

Our oven design is shown in Fig.~\ref{fig:oven}. With respect to existing ovens
such as the well-documented design of \cite{Schioppo12}, we have aimed for
further simplification and still lower power consumption by reducing the contact
surface between the oven and the surrounding parts. We use a cylindrical
stainless steel capsule, which is
resistively heated by a mineral-insulated (``Thermocoax''-type) heating
wire \footnote{ThermoExpert Deutschland GmbH; 2-wire mineral insulated heating
wire.}. This wire is welded into small grooves that were milled into the capsule
in order to improve thermal contact. The atomic beam emanates from a single
cylindrical ``capillary'' (diameter $d=0.8\,$mm, length $L=5\,$mm) drilled into
one of the bases of the cylinder. The other base is formed by a screw-on cap,
which is used to fill the oven with 2.5\,g of metallic Sr. This filling quantity
should allow non-stop operation at 450$^\circ$C for about 5 years. 

\begin{figure}[htb]
\includegraphics[width=0.9\columnwidth]{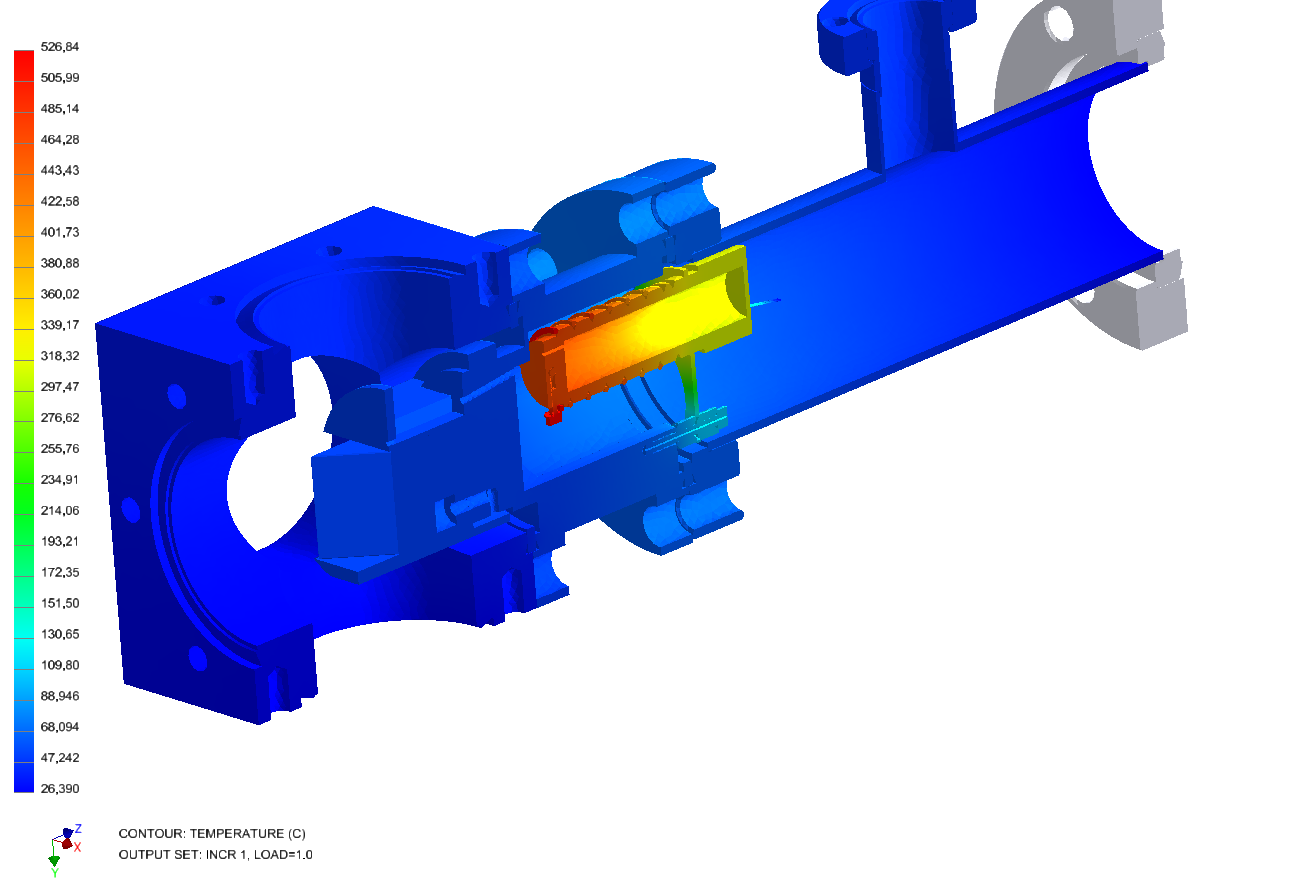}
\includegraphics[width=0.9\columnwidth]{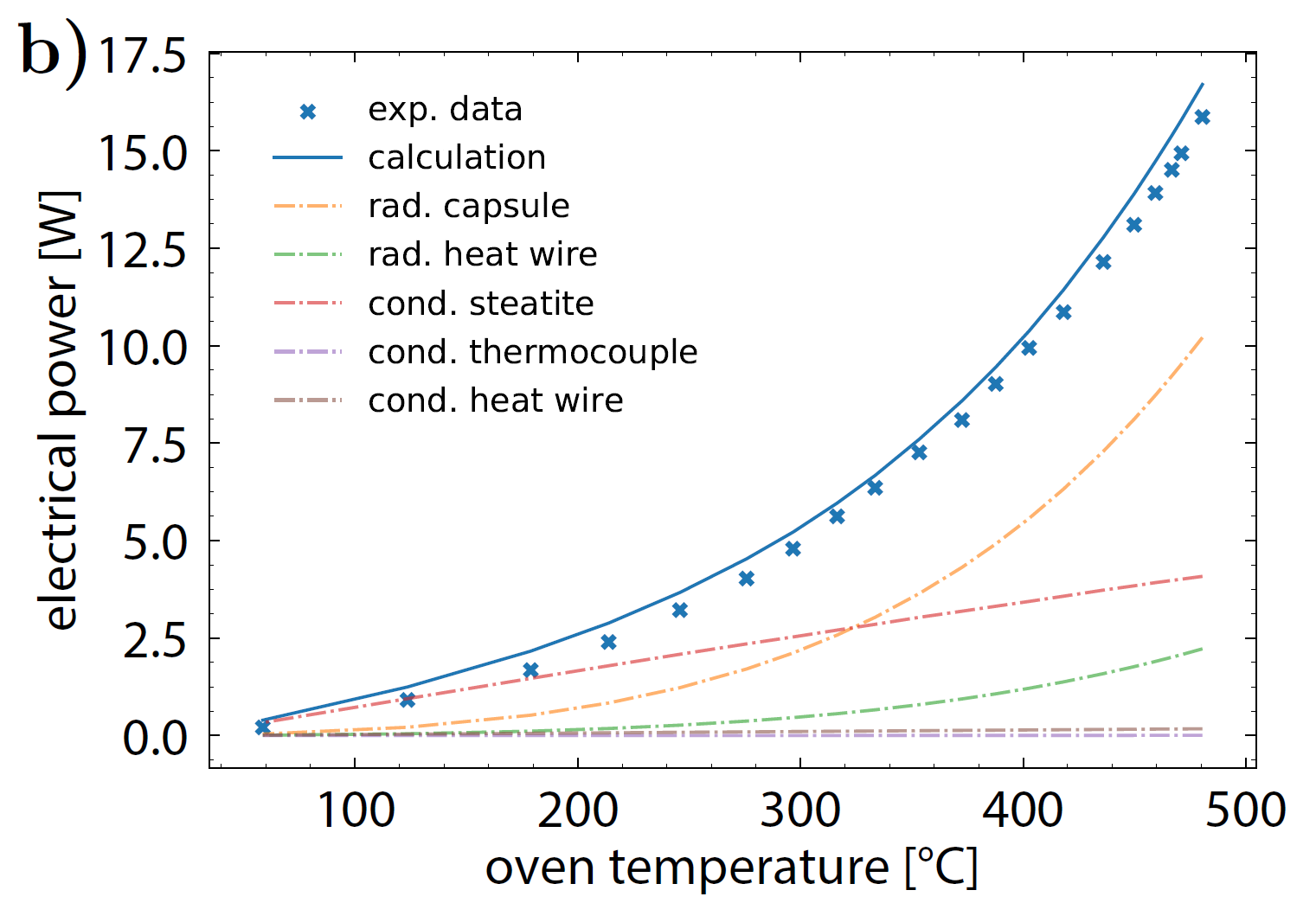}
\caption{\label{fig:oven_power}
(a) Finite-element simulation of the temperature distribution of the oven
assembly with a heating power of 15\,W. (b) Electrical power consumption vs
measured oven temperature (circles). The solid line is the result of the simple
model described in the text; dot-dashed lines indicate its individual
contributions as indicated in the legend. This calculation confirms that thermal
radiation is the leading contribution in the relevant temperature range.
}
\end{figure}

The arrangement of the heating wire aims for a slight thermal gradient along the
axis of the cylinder, with the highest temperatures on the capillary side. This was
confirmed with a finite-element simulation (Fig.~\ref{fig:oven_power}(a)). An
essential design goal is to minimize thermal conduction through the oven's
mounting structure. To reach this goal without impairing mechanical stability,
we have opted for a stainless steel holding plate of strongly reduced thickness, as visible in Fig.~\ref{fig:oven}. The plate is attached to the
surrounding 35CF flange with screws in thermally insulating steatite sleeves and spacers. 

Oven temperature is measured close to the capillary by a thermocouple with very
thin wires \footnote{}. Fig.~\ref{fig:oven_power}(b) shows the applied
electrical power vs. measured oven temperature. For an oven temperature of
450$^\circ$C, which is above our normal operating temperature (see below), the
heating power is about 13.5\,W. Remarkably, this is significantly below the
power consuption of the reference design \cite{Schioppo12} (35\,W for the same
temperature), despite the fact that that oven incorporates a heat shield while
ours does not. 
We attribute this improvement mainly to the reduced thermal conduction. This is
confirmed by a simple analytical calculation based on the heat conduction of a
metallic rod and the radiation transfer between concentric cylinders (solid and
dot-dashed lines in Fig.~\ref{fig:oven_power}(b)). Experimentally, we also find
that the temperature rise of the surrounding parts remains small in steady-state
use, requiring no water cooling.

\section{Atomic beam path}

The beam shaping goal for effusive atomic beams such as ours is to have a high
flux of atoms on axis in a solid angle corresponding to the capture region of
the magneto-optical trap that is placed further downstream. At the same time,
the flux outside that solid angle should be as low as possible, to improve oven
lifetime and to avoid problems such as covering windows or other functional
parts with a metallic layer. Many oven designs use capillary tube arrays, where
each capillary has a high length-to-diameter ratio $l_c/d_c$ with the goal of
producing good collimation, while the high number of capillaries makes up for
the low flux of a single high aspect ratio tube. To avoid clogging, however, the
capillary walls must not retain the atoms that they filter out, so they need to
be at a higher temperature than the evaporating metal (``bright walls''). Once
the oven has reached a steady state, the re-evaporated atoms make an important
and even dominant contribution to the atomic beam profile, depending on the
$l_c/d_c$ ratio. Consequently, the beam becomes broader than expected from a
naive geometrical picture. Additionally, atomic dynamics inside the capillary
are complicated and seem to defy quantitative modeling: the physicochemistry of
the wall interactions (``sticky'' and ``non-sticky'' collisions), depends on the
wall material and temperature gradient along the capillary. Furthermore, to
obtain sizeable flux, ovens often work in the transition regime from molecular
to viscous flow, which further complicates modelling. Perhaps this explains
why capillary arrays tend to perform below theoretical expectations, both in
beam intensity (atoms per second and and solid angle) and in divergence \cite{}.
In our ``bare-bones'' approach, we have therefore opted for a single capillary
with the dimensions specified in Sec.~\ref{sec:oven}. Two more diaphragms
contribute to shaping the atomic beam, as shown in Fig.~\ref{fig:vacuum}(b). The
oven section and main chamber are connected by a differential pumping hole
(length 13\,mm, diameter 1.5\,mm) which also serves for atomic beam collimation.
The beam is further collimated by a third aperture (length 2\,mm, diameter
1\,mm). The apertures form an angle of 7$^\circ$ with the horizontal plane of
the apparatus, which serves two functions. First, it directs the atomic beam to
the periphery of the MOT, rather than its center, in order to reduce the
trapped-atom losses induced by collisions with the beam. Second, it reduces the
Sr deposition rate on the 45$^\circ$ dielectric mirror which is placed at the
far end of the glass cell, and described in more detail below.

Our workshop was unable to drill the oven capillary hole under
the required $7^\circ$ angle, drilling it normal to the base surface instead.
This means that the maximum of the oven emission is stopped by the wall at the
differential pumping stage, and the atom flux in the beam is reduced. We
estimate that atom flux at a given oven temperature could be improved by at
least a factor of just by correcting this capillary angle.

\section{Assembly and vacuum}

\begin{figure}[htb]
\includegraphics[width=0.9\columnwidth]{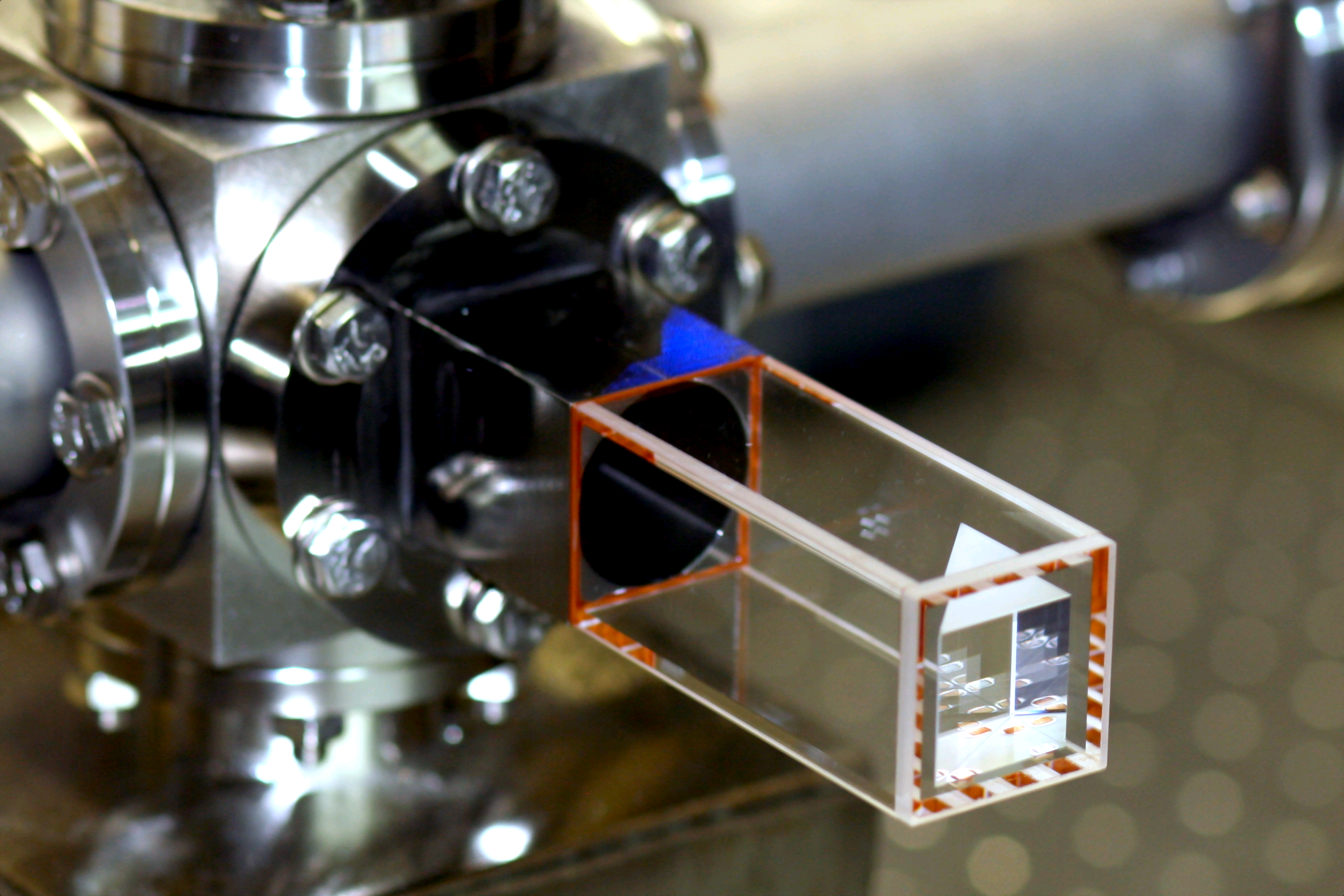}
\caption{\label{fig:cell}
Glass cell glued to titanium flange. The cell is a low-cost catalog part
intended for spectroscopy applications. The custom machined titanium flange has
a coefficient of thermal expansion matching that of the glass, enabling bakeout
with limited mechanical stress at the interface. The cell-flange joint is a
low-outgassing epoxy glue.}
\end{figure}

In the compact vacuum apparatus shown in Fig.~\ref{fig:vacuum}, one key element
is the glass cell glued to a titanium flange. Titanium has a thermal expansion
coefficient of approximately $8.6\dip{-6}\,\mbox{K}^{-1}$, closely matching that
of BK7 and many other crown glasses. Such glasses are used in many commercial
spectroscopy cells, offering high optical quality at low cost. Our apparatus
currently uses an uncoated standard cell\footnote{Hellma 402.013 OG} with outer
dimensions $70\,\mbox{mm}\times 30\,\mbox{mm}\times 30\,\mbox{mm}$ with a wall
thickness of 2.5\,mm.  The cell is glued to
the flange using a low-outgassing, filled epoxy glue \footnote{Epo-Tek H77T} to
the titanium flange \footnote{Grade 5, 6Al-4V}. Inside the cell, the dielectric
mirror-coated prism is glued \footnote{Epo-Tek 353ND} onto the bottom surface
using three quartz glass spacers with dimensions $2.5\,\mbox{mm}\times
2.5\,\mbox{mm}\times 1\,\mbox{mm}$. Fig.~\ref{fig:cell} shows the assembled
cell.

The oven was filled under a protective Ar atmosphere using a glove bag, then
screwed onto the vacuum apparatus and pumped down immediately with a
turbomolecular pump through the 35\,mm angle valve visible in
Fig.~\ref{fig:vacuum}. In normal operation after the bakeout phase, the oven
section is pumped by a $3\mbox{l}/\mbox{s}$ ion pump, and the main chamber by a
$25\mbox{l}/\mbox{s}$ ion pump and a non-evaporative getter\footnote{SAES
Getters, CapaciTorr D 200} which we have not activated yet. Some of the vacuum
components have been pre-baked in a vacuum oven at $250^\circ$C for 48\,h before
the the vacuum system was assembled. The complete system was baked at
$150^\circ$C for 48\,h. This low temperature was chosen as a precaution for the
glued cell\footnote{Higher temperatures up to about 180$^\circ$C should be
possible, but we did not investigate this because of the envisaged use with a
high-finesse microcavity, which will also limit the bakeout temperature to
$150^\circ$C.}. This minimal baking schedule resulted in a room-temperature base
pressure of $1\times 10^{-7}\,$mbar in the oven section and $1.3\times 10^{-10}$
in the main chamber. The oven base pressure is much higher than expected, but we
did not investigate this further as its contribution to the main chamber
pressure remains small due to the differential pumping hole. At an oven
temperature of $450^\circ$C, the pressures rise to $2\times 10^{-7}$ in the oven
section and $5\times 10^{-10}$ in the main chamber.  The latter pressure will
further decrease once the getter is activated.

\section{Zeeman slower and MOT coils}

As the mean longitudinal speed in the atomic beam is hundreds of meters per
second, much higher than the MOT capture range, beam slowing is mandatory for
efficient MOT loading. The strong blue transition in Sr enables high
deceleration, so that the beam can be slowed to near zero mean longitudinal
speed on a very short distance on the order of 15\,cm. Traditional Zeeman
slowers have specially engineered coils that attempt to exactly compensate the
mean Doppler shift by the local Zeeman shift at every point along the slowing
path. However, orders of magnitude are such that a good approximation to this
ideal magnetic field profile can be achieved by the decaying field of the MOT
coils by appropriately sizing and orienting them. This is the approach we have
followed here.

Zeeman slowing reduces only the longitudinal velocity. The transverse velocity
spread is initially determined by the oven nozzle and beam apertures, and
increases due to spontaneous emission during Zeeman slowing. Therefore, a long
oven-MOT distance can significantly reduce the atomic flux into the MOT capture
region, even when transverse laser cooling is applied to the beam. In our setup,
the glued cell enables a total beam path length as short as the theoretical
stopping distance (cf.~Fig.~\ref{fig:vacuum}(b)). This short distance is also
required for using the MOT magnetic field for Zeeman slowing. 

\begin{figure}[htb]
\includegraphics[width=0.8\columnwidth]{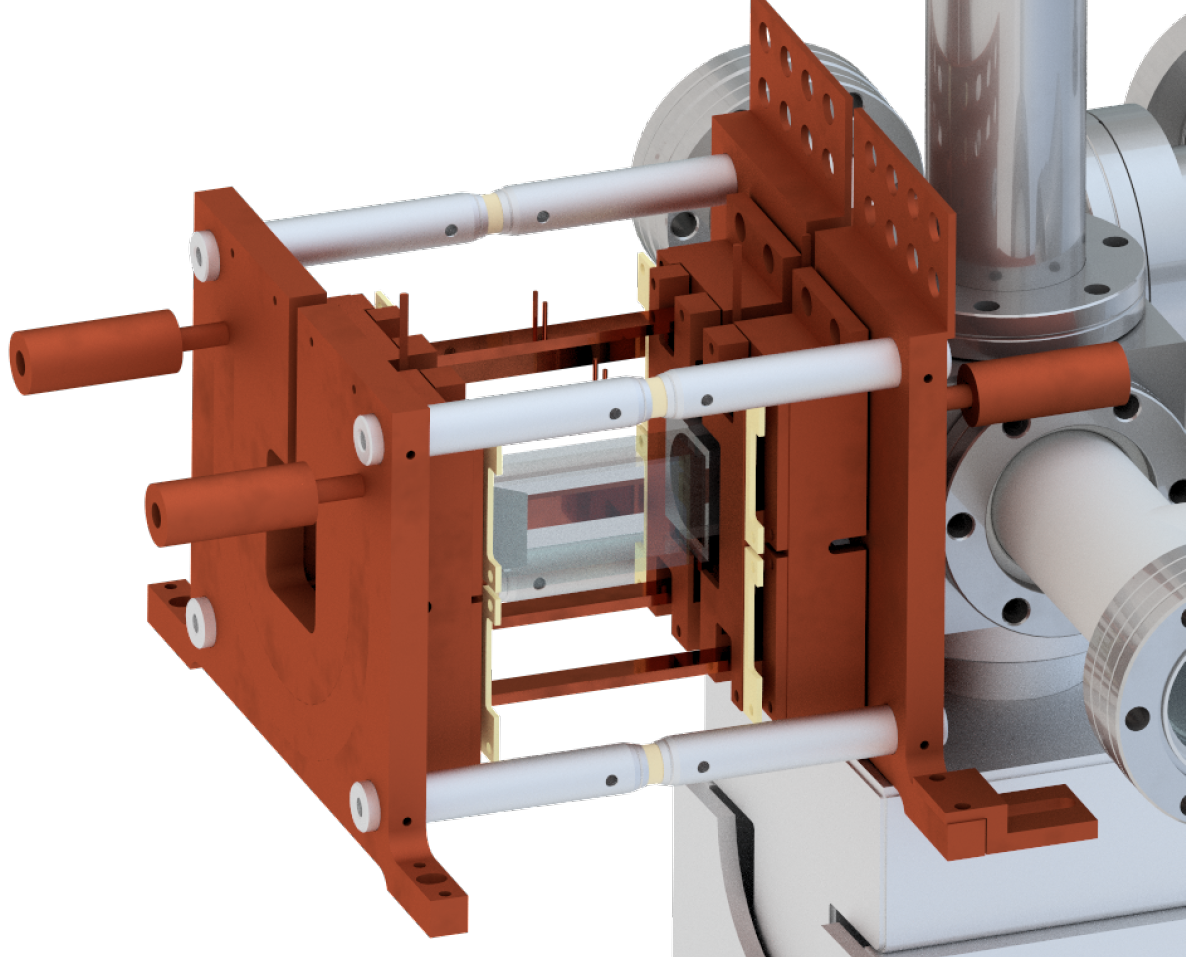}\\
\vspace{\baselineskip}
\includegraphics[width=0.8\columnwidth]{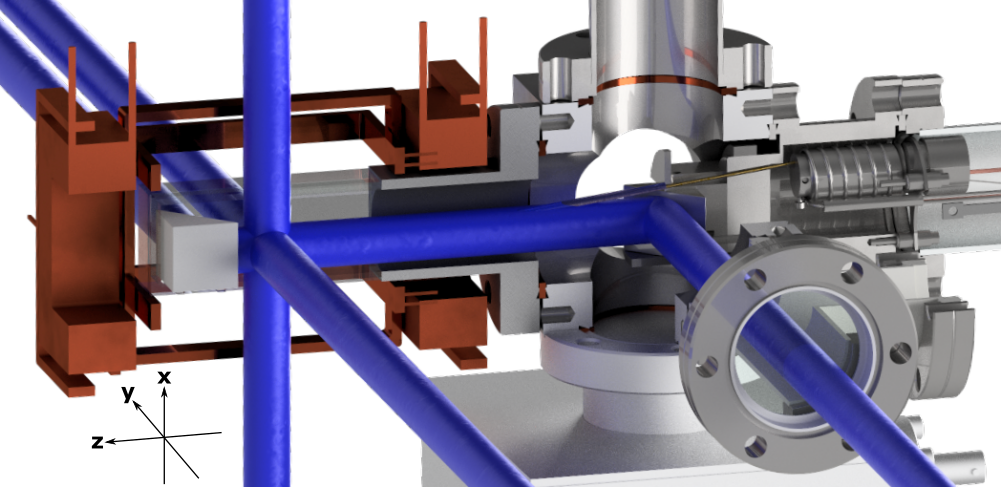}
\caption{\label{fig:MOTcoils} (a) Magnetic coil assembly. The MOT coils have 60 turns each and
also produce the magnetic field for the Zeeman slower. These coils are embedded
in a slotted copper structure containing water circulation channels for heat
removal. 3 pairs of bias field coils have 11 turns each. All coils are
self-supporting and wound from solid copper wire \footnote{Imetron GmbH, Umkirch,
Germany.}. (b) Beam path of the blue MOT beams. }
\end{figure}

Fig.~\ref{fig:MOTcoils} shows the magnetic coil arrangement and the MOT and
Zeeman slowing laser beams. The quadrupole axis is horizontal, aligned with the
main axis of the apparatus. The MOT coils, 60 turns each, are square with inner and outer
dimensions of $40\,\mbox{mm}\times 40\,\mbox{mm}$ and $70\,\mbox{mm}\times
70\,\mbox{mm}$ respectively. These dimensions, with a distance larger than the
usual "anti-Helmholtz" condition, are chosen to produce suitable Zeeman slowing
and MOT fields simultaneously. At their nominal current of $I_0=25\,$A, the
on-axis magnetic field gradient in the MOT center is approximately
$33\,\mbox{G}\mbox{cm}^{-1}A^{-1}$, in good agreement between calculation and
measurement with a fluxgate probe. Each coil dissipates
approximately 35\,W, requiring water cooling. Therefore, the MOT coils as well
as three pairs of bias field coils (11 turns each) are mounted on a common copper support
structure, which is slotted to reduce Foucault currents. The structure
incorporates water circulation to cool the assembly with an external chiller.

\section{Laser system}
Like other alkaline-earth atoms, strontium has a strong $^1 S_0-{}^1 P_1$
dipole-allowed transition. Its center wavelength is $\sim 461\,$nm and its
linewidth $\Gamma= 30.5\,$MHz (``blue'' transition). This transition allows
fast cooling, but with a high Doppler limit of about $0.7\,$mK due to the large linewidth.
Therefore, experiments add a second MOT stage using a much narrower $^1 S_0-{}^3
P_1$ transition (689\,nm, $\Gamma= 7.4\,$kHz, ``red'' transition). Additionally,
repumping lasers are needed to recover atoms accumulating in the $^3P_2$ state.
Several options exist, we have chosen the ``classical'' 707 and 679\,nm
wavelengths.

Aiming for simplicity and low cost, we have employed neither second-harmoic
generation nor tapered amplifiers. As described in the following, our laser
system uses external-cavity diode lasers (ECDLs) throughout, and employs
injected slave diodes where higher power is required. 

\subsection{Blue laser system}
\begin{figure}[htb]
\includegraphics[width=0.9\columnwidth]{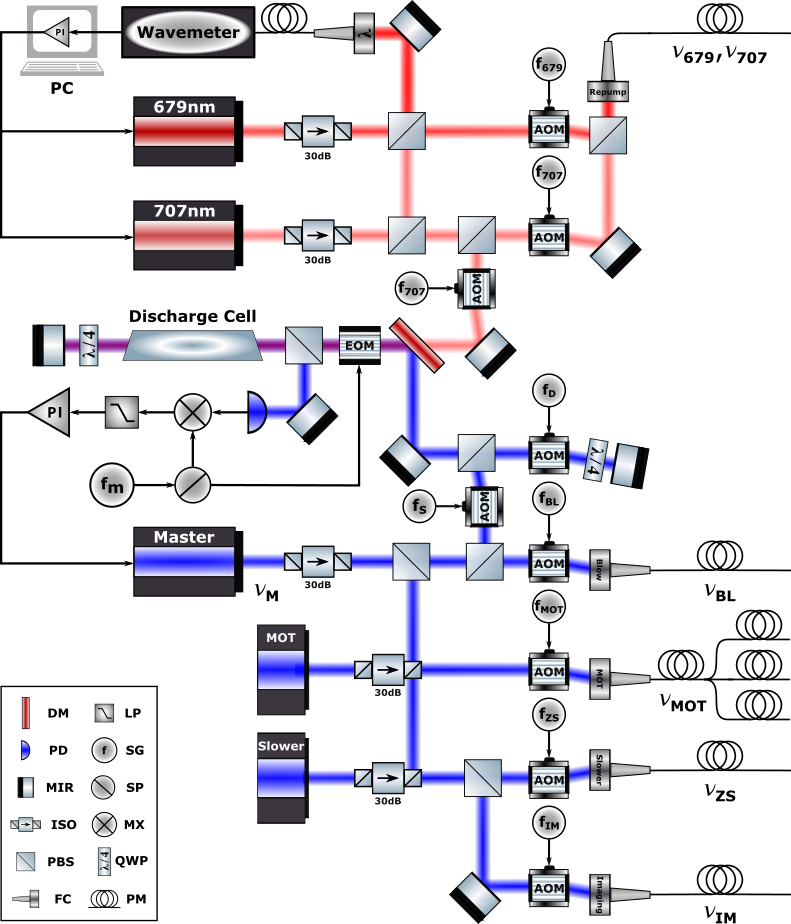}
\caption{\label{fig:bluelasers}
Simplified schematic of the blue laser system. Half-wave plates have been
omitted for simplicity. DM: dichroic mirror, PD: photodiode, MIR: mirror, ISO: optical isolator, PBS: polarizing beam splitter, LP: low-pass filter, SG: signal generator, SP: RF splitter, MX: RF mixer, FC: fiber coupler, PM: polarization-maintaining fiber, AOM: acousto-optic modulator, EOM: electro-optic modulator. 
}
\end{figure}

Fig.~\ref{fig:bluelasers} shows the blue laser system. The master laser is a
commercial external cavity diode laser (ECDL) \footnote{MOGLabs LDL; Nichia
NDBA116T diode} in Littrow configuration. It is stabilized to the Doppler-free
absorption signal of the $^1S_0 \to ^1P_1$ transition, then amplified by
injection locking two slave diodes \footnote{Nichia NDB4216E}, and further used
as probe. Instead of a heated spectroscopy cell, we use a hollow cathode
discharge cell \footnote{16Hamamatsu L2783-38NE-SR}, which is powered by a
low-noise high-voltage DC-DC converter \footnote{ISEG CPp 05 206 24 05 module}.
The slave laser frequencies are shifted to the required detunings by means of
single-pass acousto-optical modulators (AOMs), which also serve as fast switches.
The first slave laser is shifted by $-30\,$MHz and delivers the optical power
for the 3D MOT. Its beam is divided by a commercial fiber splitter
\footnote{Evanescent Optics; spliceless PM coupler 1x3 array} before the light
is sent to the main optical table. The second slave laser serves as cooling
laser for the Zeeman slower and is shifted by $-550\,$MHz. We also tested
injection locking a low-cost multi-mode, high-power laser diode \cite{Pagett16}, 
but no appropriate model compatible with the 5.6mm laser diode
socket setup had been found as their emission spectrum merely touched the
required wavelength of 461nm.

\subsection{Repumpers}

Among several repumping transitions that have been demonstrated, we have chosen
the widespread repumping scheme on the $(5s5p)^3P_{0,2} \to (5s6s)^3S_1$
transitions, which occur at wavelengths of 679.289\,nm and 707.202\,nm
respectively. Diode lasers with ample power are readily available for both
wavelengths, we use ``cateye'' ECDLs \footnote{MOGLabs CEL002}.. Additionally,
these lasers may also be useful tools by themselves, such as for state-selective
imaging \cite{Cooper18,Bothwell19,Madjarov20}.
 While the 707\,nm  repumper can be stabilized on the same hollow cathode lamp as the 461\,nm laser \cite{Norcia16}, using the same approach for the 679\,nm laser is challenging due to a weak signal-to-noise ratio. Therefore, we use a microcontroller-based digital lock to a wavemeter \footnote{HighFinesse WS7-60}. This scheme is easily extended to the 707\,nm as well and we employ it for both repumpers.

\subsection{Red laser system}
\begin{figure}[htb]
\includegraphics[width=0.9\columnwidth]{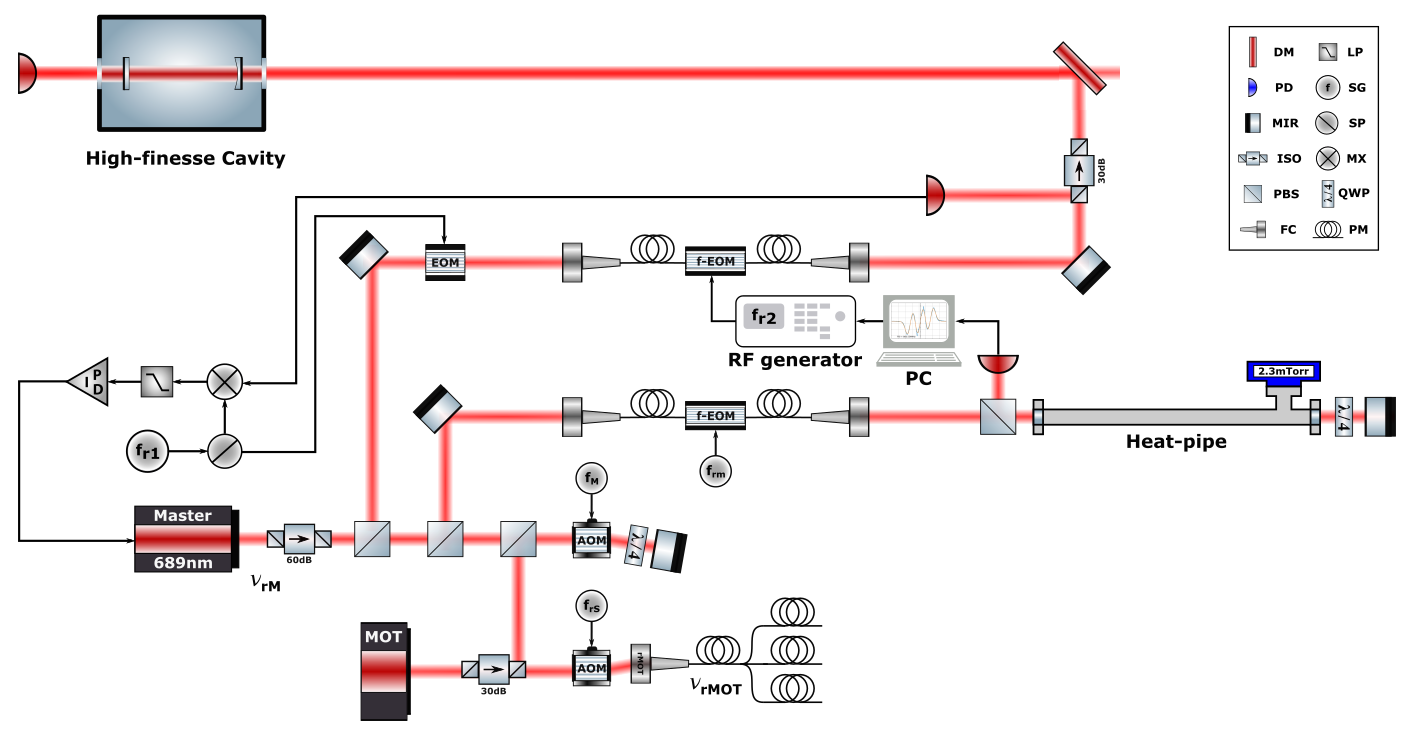}
\caption{\label{fig:redlasers}
Simplified schematic of the red laser system. DM: dichroic mirror, PD: photodiode, MIR: mirror, ISO: optical isolator, PBS: polarizing beam splitter, LP: low-pass filter, SG: signal generator, SP: RF Splitter, MX: RF mixer, FC: fiber coupler, PM: polarization-maintaining fiber, AOM: acousto-optic modulator, EOM: electro-optic modulator, f-EOM: fiber EOM.}
\end{figure}

The architecture of the red laser system, shown in Fig.~\ref{fig:redlasers},
takes into account a future use of the 698\,nm clock transition. The master
laser for the red MOT is a ``cateye'' ECDL\footnote{MogLabs CEL002} which is
locked onto the same ultrastable reference cavity\footnote{Stable Laser Systems
Inc, free spectral range: 1.5\,GHz, measured finesse: 33000 at 689\,nm.} that
also serves for the 698\,nm laser. A Pound-Drever Hall lock stabilizes the laser
to the cavity resonance. From the error signal, we estimate \cite{Domenico10} a
sub-100\,Hz linewidth (63.8\,Hz closed-loop area, 27.5\,Hz closed-loop flicker).
A home-made heat pipe spectroscopy cell provides absolute frequency calibration.
The cavity drift, measured with this cell, is $5.94(3)\,\mbox{kHz}/\mbox{d}$,
which is low enough so that one calibration of the atom-cavity detuning per day
is sufficient for operating the red MOT without loss of capture efficiency. A
double-pass AOM provides frequency modulation, in particular for the SWAP MOT
(cf.~\ref{sec:expRed}). Fig.~\ref{fig:redlasers} gives further details.

\subsection{MOT and Zeeman slowing beams}

Our setup uses two optical tables. Polarization-maintaining fibers are used to transfer the blue and red beams from the laser table to the main table. MOT beams are slightly focussed to compensate for the transmission losses of the uncoated cell. For efficient use of the red laser power, our red MOT beams have a relatively small $1/e^2$diameter of 4.8\,mm. 

To align the red MOT beams to the blue
MOT position, we periodically switch the magnetic gradient coils with a period
in the 100\,ms range to make the blue MOT blink. Its fluorescence is observed on
a photodiode while the red beams are frequency-scanned across resonance. The
alignment goal is to deplete the blue fluorescence as much as possible, for each
red beam axis.

\section{Experimental results: Atomic beam}
The atoms leave the effusive oven with an average velocity of approximately
500\,m$/$s. We have first characterized the oven alone in a test chamber to
verify its function. Once inserted into the complete apparatus described above,
we have measured the properties of the collimated beam arriving in the glass
cell using a transverse probe beam of waist $w_0 \approx 140\,\mu$m and optical
power $P0\approx 55\,\mu$W, resulting in a saturation parameter $s_0\approx
4.3$. Using translation stages in the longitudinal and transverse directions, we
have taken scanned-frequency spectra at several positions. Each spectrum was
fitted with the expected spectrum taking into account all Sr isotopes, with the
density and transverse velocity as free parameters. We find a small transverse
velocity of $7(2)\,\mbox{m}\mbox{s}^{-1}$, compatible with the expected velocity
from the initial Gaussian distribution and the diameter and distance of the
collimation hole. From the position-resolved measurements, we find a beam
divergence (HWHM angle) of $\delta\theta= 1.2(1)^\circ$, in agreement with the
expectation from the mechanical parameters. The resulting total flow rate is
$3.0(2)\cdot 10^{10}\mbox{at}\mbox{s}^{-1}$ at $460^\circ$C.

\section{Experimental results: Loading and blue MOT characteristics}
We have used the setup to cool and trap the most abundant, bosonic isotope
$^{88}$Sr, as well as the fermionic $^{87}$Sr. For the MOT characterization
below, we use $^{88}Sr$. Following established practice, atoms from the slowed
beam are first cooled and trapped using the blue transition. The 679\,nm and
707\,nm repumpers are both switched on throughout the MOT loading. Trapped atom
characteristics are measured using absorption imaging on this same transition,
using an imaging beam in the horizontal plane, forming an angle with the
horizontal MOT beam axis. Additionally, MOT fluorescence is collected by a lens
and recorded with a photodiode.

\begin{figure}[htb]
\includegraphics[width=0.9\columnwidth]{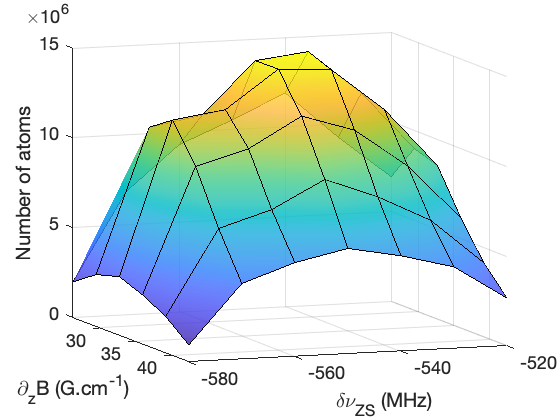}\\
\vspace{\baselineskip}
\includegraphics[width=0.9\columnwidth]{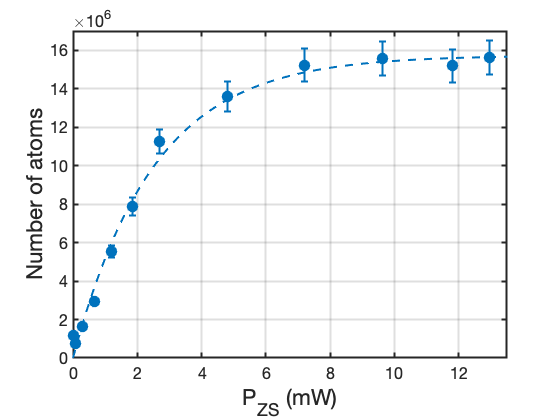}
\caption{\label{fig:MOTscan} (a) Number of atoms in the blue MOT, $N_b$, after
3\,s loading as a function of MOT coil current (shown is the longitudinal
magnetic gradient) and detuning of the Zeeman slowing beam. Other parameters
are: Zeeman slower beam power $P_{ZS}=12\,$mW, blue MOT beam power $P_b=11\,$mW
(before the glass cell) and detuning $\delta_b=-32\,$MHz. (b) $N_b$ as a
function of the Zeeman beam power (measured before the glass cell). }
\end{figure}

Due to the double use of the magnetic gradient coils, Zeeman slowing and MOT
loading are intrinsically coupled by the gradient coil current $I_g$. To
approach the optimum parameters, we have performed a two-dimensional scan of
$I_g$ (shown as the on-axis magnetic gradient $\partial_z B$) and the detuning
of the Zeeman slowing beam, $\delta\nu_{ZS}$, with all other parameters kept
constant. The result is shown in Fig.~\ref{fig:MOTscan}(a). A broad maximum
appears around $\partial_z B=32.5\,\mbox{G}/{cm}$ and
$\Delta\nu_{ZS}=-545\,$MHz, with a trapped atom number $N_b\approx 1.5\dip{7}$.
The only other adjustable parameter of the Zeeman slower is the beam power
$P_{ZS}$, which is scanned in Fig.~\ref{fig:MOTscan}(b). The atom number
increases linearly for low powers and reaches a plateau for $P_{ZS}\gtrsim
8\,$mW. In the following, we always use powers in the saturated regime,
$P_{ZS}\ge 10\,$mW.

\begin{figure}[h]
\includegraphics[width=0.32\columnwidth]{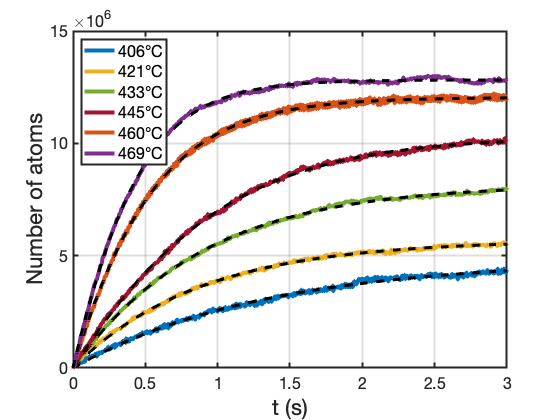}
\includegraphics[width=0.32\columnwidth]{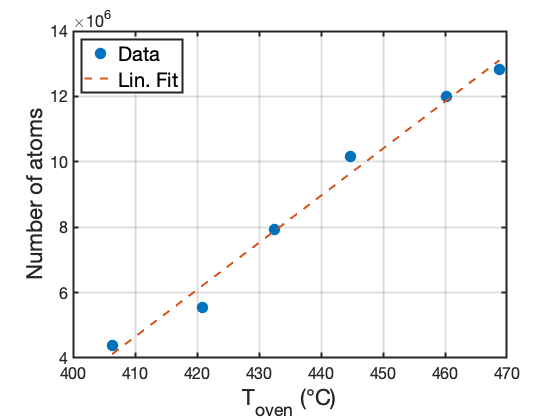}
\includegraphics[width=0.32\columnwidth]{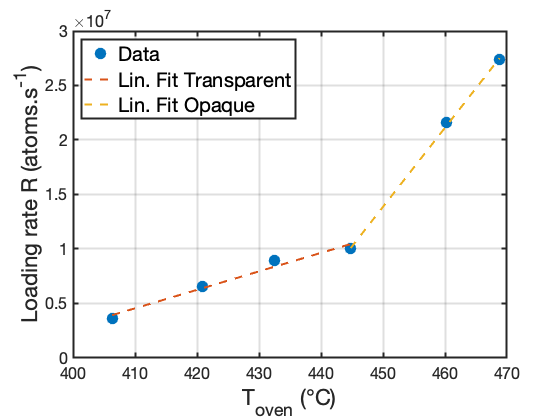}
\caption{\label{fig:MOTloading}
(a) MOT loading curves, measured in fluorescence, for different (unevenly
spaced) oven temperatures. (b) Steady-state atom number as a function of oven
temperature. No clear saturation effect is visible for the temperatures
explored. (c) Loading rate $R$ extracted from a fit to the loading curves,
suggesting two regimes with a transition around $450^\circ$C. }
\end{figure}

In a beam-loaded MOT, higher atomic beam flux increases the capture rate, but
also induces additional atom loss (in addition to collisions with the background
gas), which reduces trap lifetime and will eventually limit the steady-state
atom number when this loss term dominates. Fig.~\ref{fig:MOTloading}(a) shows
MOT loading curves, measured in fluorescence, for different
oven temperatures. These curves are obtained by switching on the Zeeman slowing
beam at $t=0$. For
$T>460^\circ$C, a loading time of 1.5\,s is sufficient to reach the steady
state. Plotting the steady-state atom number as a function of oven temperature
(Fig.~\ref{fig:MOTloading}(b)), no clear saturation effect is visible for the
temperatures explored. Indeed, fitting the loading and decay curves with the
canonical rate equation model, the resulting one-body loss rate increases by
less than a factor of two over the background value of $1\,\mbox{s}^{-1}$.
Interestingly, the loading rate $R$ extracted from the same fit
(Fig.~\ref{fig:MOTloading}(c)) suggests two regimes with a transition around
$T_c\sim 450^\circ$C. This is the temperature range where the mean free path in
the atomic vapor becomes comparable to the transverse dimensions of the nozzle
and diaphragms, i.e., where the beam transitions from the transparent to the
``opaque'' regime. In order to keep a typical MOT lifetime $\ge 1\,$s and yet
have a sufficient number of atoms in the blue MOT, we operate the oven at $T_c$
in the following.

Varying the detuning of the blue MOT beams and measuring temperature in time of
flight, we find a broad plateau of $T\sim 10\,$mK, significantly above the
Doppler limit, for detunings $-30\,\mbox{MHz}\lesssim\delta \nu_{b}\lesssim
-7\,\mbox{MHz}$. In order to further reduce this temperature and increase
density before transfer to the red MOT, we have implemented a compression
sequence where the $P_{ZS}$ is switched off, $P_b$ is reduced to 200\,$\mu$W per
beam in 100\,ms, and simultaneously, the MOT gradient is increased to
$\partial_z B=65\,\mbox{G}/\mbox{cm}$. This results in a sample of typically
$N=2\dip{7}$ atoms at a temperature $T=2.4\,$mK. Whith less extensive
optimization for $^{87}$Sr, we currently obtain $N=9\dip{5}$ at $T=1.2\,$mK for
that isotope.

\section{Experimental results: Red MOT}
\label{sec:expRed}

At the end of the blue MOT stage, the Doppler width $\sigma=\sqrt{k_B
T/(m\lambda^2)}$ is on the order of 1\,MHz for a $T=2\,$mK cloud. In order to
load these atoms into the red MOT, where the transition linewidth is
$\Gamma_r=7.4\,$kHz, a broadening mechanism is required. This can be
accomplished by modulating the frequency of the red MOT laser, either
sinusoidally \cite{Katori99} or with a sawtooth function (``SWAP'' MOT)
\cite{Norcia18,Muniz18,Snigirev19}. Here, we opt for the SWAP MOT. 

Due to the low initial velocity spread and the narrow linewidth of the red
transition, appropriate magnetic field gradients for the red MOT are much lower
than for the blue one. $200\,\mu$s after switching off the blue MOT, we use a
solid-state relay-based circuit switch to lower the magnetic field gradient by a
factor $\sim 50$ to $\partial_z B=1.3\,\mbox{G}/\mbox{cm}$ with an
exponential-decay time constant $160\,\mu$s. We experimentally observe that
switching on the red MOT beams 5\,ms before the gradient switching increases the
recapture fraction; this 5\,ms overlap is used in all the data below.

\begin{figure}[htb]
\includegraphics[width=\columnwidth]{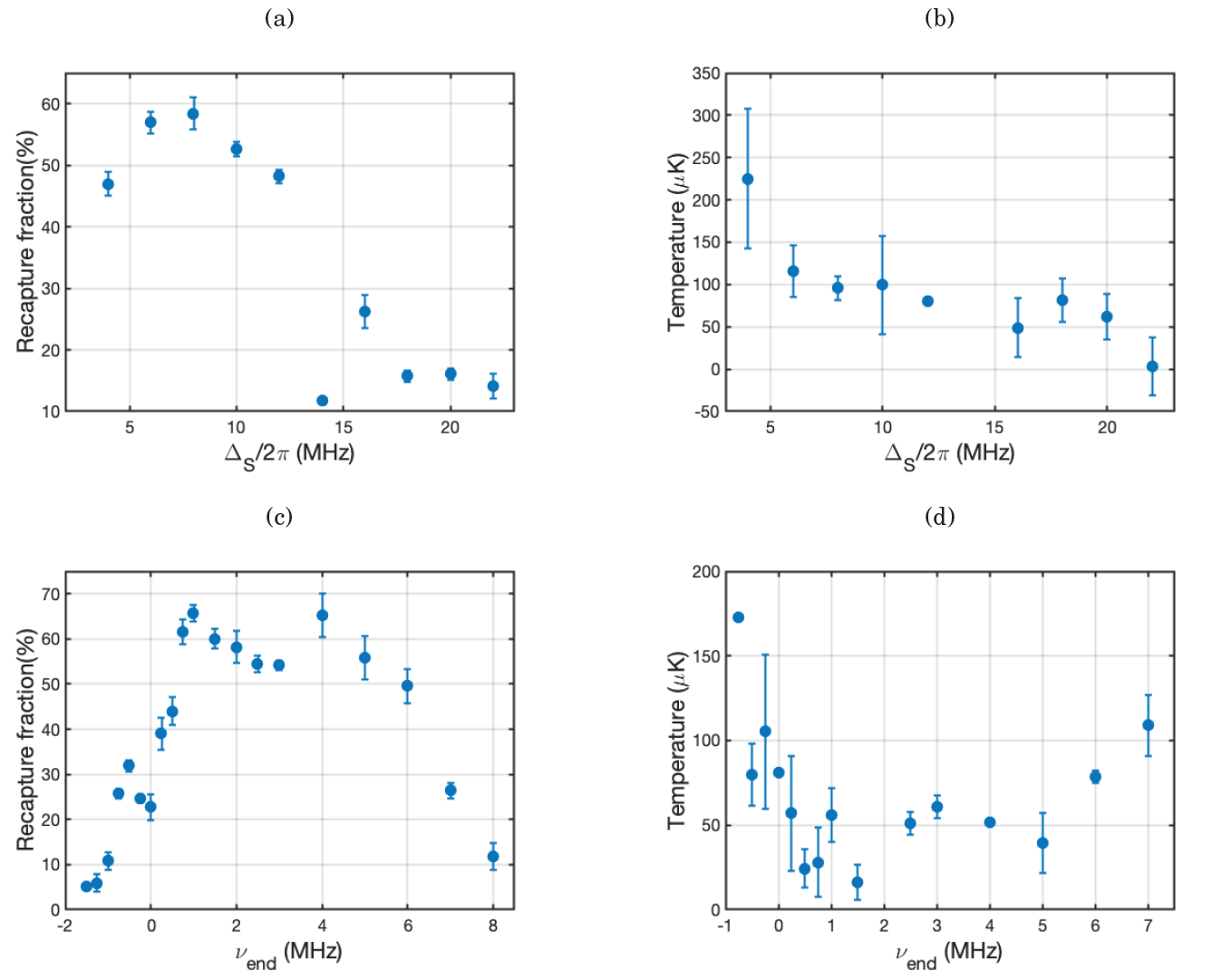}
\caption{\label{fig:SWAP}
(a) Recapture fraction from the blue MOT and (b) temperature estimated from the
central Gaussian in time-of-flight absorption images, for varying SWAP amplitude
$\Delta_S$ with a fixed final detuning $\nu_\text{end}=1.0\,$MHz. (c) Recapture
and (d) temperature as a function of $\nu_\text{end}$ with a fixed amplitude
$\Delta_S=8.0\,$MHz. Both datasets are taken at the fixed red beam power of
6.2\,mW per axis and for a number of sweeps  $N_{SW}=600$.}
\end{figure}

For the period $T_{ST}$ of the sawtooth frequency modulation, a previous study
\cite{Snigirev19} has shown that the atom number in the blue MOT improves for
shorter $T_{ST}$ until a plateau is reached for $T_{ST}\sim 2\tau_b$, where
$\tau_b\sim 21.3\,$\textmu s is the lifetime of the $^3 P_1$ excited state. In
the following, we have used $T_{ST}=50\,\mbox{\textmu s}\approx 2\tau_b$. Two
more parameters determine the modulation: its amplitude $\Delta_S$, and its
final detuning $\nu_{\text{end}}$. We have studied the influence of these two
parameters using a test sequence with $N_{SW}=600$ sweep periods (i.e.,
$t_{\text{SWAP}}=30\,$ms) and a fixed red laser power of 6.2\,mW per axis
(Fig.~\ref{fig:SWAP}). Using a fixed $\nu_{\text{end}}=1.0\,$MHz, and varying
$\Delta_S$, we observe that the number of atoms recaptured from the blue MOT is
maximized in the region 6\,MHz$\le\Delta_S\le 8\,$MHz, peaking at $\sim 60\%$ in
this sequence, followed by a drop for higher modulation amplitudes
(Fig.~\ref{fig:SWAP}(a)). This drop can be explained by the reduced intensity
available per frequency band. Nonetheless, atoms are still trapped for
amplitudes as high as 25\,MHz, proving the robustness of the technique. The
corresponding temperatures (Fig.~\ref{fig:SWAP}(b)) are estimated from
absorption images, showing a drop from $T>200\,$\textmu K to $T<100\,$\textmu K
around $\Delta_S=5\,$MHz, followed by a slower decrease to $T\lesssim
50\,$\textmu K for higher $\Delta_S$ up to 20\,MHz.Therefore, the optimum
$\Delta_S$ is a compromise between temperature and atom number. Fixing
$\Delta_S=8\,$MHz and varying $\nu_{\text{end}}$ from $-2\,$MHz to $+8\,$MHz,
the recaptured fraction reaches a plateau for $1\lesssim\nu_{\text{end}}\lesssim
4\,$MHz (Fig.~\ref{fig:SWAP}(c)), and the temperature 
has a minimum $T\sim 50\,$\textmu K in the same region (Fig.~\ref{fig:SWAP}(d)). 

\begin{figure}[htb]
\includegraphics[width=0.49\columnwidth]{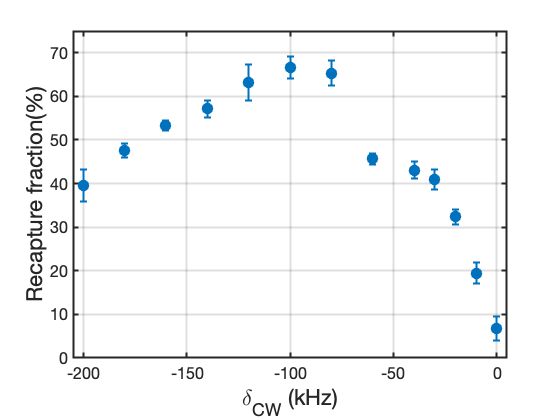}
\includegraphics[width=0.49\columnwidth]{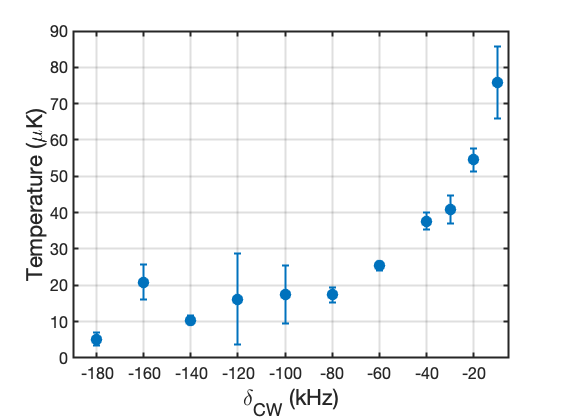}
\caption{\label{fig:redMOT}
}
\end{figure}

We have also studied the influence of the power $P_b$ of the red MOT beams and
found that ramping down $P_b$ during the SWAP phase allows reaching lower
temperatures without significant atom loss, compared to constant $P_b$. Still,
these temperatures remain in the $T\sim 50\,$\textmu K range, much higher than
the Doppler limit. Therefore, we have added a second red MOT phase where the
detuning is kept constant while the intensity is ramped down to low values
within another 50\,ms. This leads to temperatures approaching the recoil limit,
which we cannot measure precisely with our current low-budget CCD camera.
Indeed, the camera's full-well capacity limits imaging pulses to
low photon numbers, and combines with the readout noise to make long time of
flights at low atom numbers hard or impossible. At the short TOFs that we can
use, the non-Gaussian cloud shape leads to large systematic errors. Therefore,
the temperatures in Fig.~\ref{fig:redMOT} only indicate trends below 20\,\textmu K.
Nevertheless, these results show that a recapture fraction of
65\,\% of the blue MOT atoms remains at the end of a combined SWAP and
constant-detuning red MOT phase, and reaches minimum temperatures below
$20\,$\textmu K and probably lower, at a detuning $\delta_{\text{CW}}\sim
-100\,$kHz.

In an application such as cavity QED, where the target atom number is orders of
magnitude lower than the $\sim 10^7$ atoms obtained here, the temperature could
be further reduced at the expense of atom number, for example by adding a
molasses phase.

\section{Conclusion and outlook}
We have presented a compact apparatus for producing ultracold strontium in which each major subsystem has been deliberately simplified. Since the design principles rest on general features of the alkaline-earth level structure rather than on peculiarities of Sr, most of the concepts should carry over to other alkaline-earth and alkaline-earth-like species with only moderate adaptation. Besides lowering the entrance barrier into the alkaline-earth field, we hope that the design presented here will contribute to progress in the field because of the increased reliability that results from reduced complexity. In addition to portable metrology systems \cite{Grotti18,Bothwell25}, this may benefit any alkaline earth-like project which does not require extremely high atom flux. Here, reduced complexity in the atom production part means freedom to add complexity elsewhere. One example is the combination of arrays of individually controlled alkaline-earth atoms in optical tweezers with a cavity quantum electrodynamics platform, a promising direction, currently witnessing first experimental realizations with alkali atoms \cite{Ye26,Roschinski26,Picot26}. The narrow optical transitions of earth alkali-like atoms provide additional incentive to realize such system, for quantum simulations and readout in quantum processors, but also for multi-parameter quantum metrology \cite{Baamara23}.

\section*{Acknowledgments}
\label{sec:ack}
We thank Guglielmo Tino, Ekkehard Peik and Tanja Mehlst\"aubler for valuable
advice in the early phase of the project, and acknowledge insightful discussions
on Zeeman slower design with Edward G. Marti and Sebastian Blatt. This work is
supported by the European Research Council (ERC) (Advanced Grant 671133 EQUEMI).



\bibliographystyle{apsrev4-2}
\bibliography{JRLibrary}

\end{document}